\documentstyle[12pt]{article}

\renewcommand{\baselinestretch}{1.65}

\newtheorem{theorem}{Theorem}[section]
\newtheorem{lemma}{Lemma}[section]

\newfont{\Mb}{msbm10}
\newcommand{\R}{\mbox{\Mb\symbol{82}}}
\newcommand{\C}{\mbox{\Mb\symbol{67}}}

\begin{document}

\renewcommand{\baselinestretch}{1.1}

\title{\Large \bf On the Algebraic Structure of Second
                  Order Symmetric Tensors in 5-Dimensional Space-times \\ }

\author{
G.S. Hall\thanks{
{\sc internet: gsh@maths.abdn.ac.uk}  }, \ \
M.J. Rebou\c{c}as\thanks{
{\sc internet: reboucas@cat.cbpf.br}  },  \ \
J. Santos\thanks{ {\sc internet: janilo@dfte.ufrn.br} } \ \   and  \ \
A.F.F. Teixeira \thanks{ {\sc internet: teixeira@cat.cbpf.br} } \\
\\  \vspace{1mm}
$^{\ast}~$Department of Mathematical Sciences \\
          University of Aberdeen \\ Aberdeen AB9 2TY,
          Scotland -- U.K.  \\
\\
$~^{\dagger}~^{\S}~$Centro Brasileiro de Pesquisas F\'\i sicas \\
                    Departamento de Relatividade e Part\'\i culas \\
                    Rua Dr.\ Xavier Sigaud 150 \\
                    22290-180 Rio de Janeiro -- RJ, Brazil \\
\\
$^{\ddagger}~$Universidade Federal do Rio G. do Norte \\
             Departamento de F\'{\i}sica, Caixa Postal 1641 \\
             59072-970 Natal --   RN,  Brazil \vspace{3mm}  \\
        }

\date{\today}

\maketitle

\vspace{2mm}

\begin{abstract} \vspace{2mm}
A new approach to the algebraic classification of second order
symmetric tensors in 5-dimensional space-times
is presented. The possible Segre types for a symmetric two-tensor
are found. A set of canonical forms for each Segre type is obtained.
A theorem which collects together some basic results on the
algebraic structure of the Ricci tensor in 5-dimensional space-times
is also stated.
\end{abstract}

{\raggedright
\section{Introduction} }   \label{intro}
\setcounter{equation}{0}

In general relativity, it is well known that the curvature tensor
can be uniquely decomposed into three irreducible parts, namely
the Weyl tensor, the traceless Ricci tensor and the Ricci scalar.
Petrov and others~\cite{petrov1}~--~\cite{penrose2} have discussed the
algebraic classification of the Weyl part of the curvature tensor.
The  Petrov classification as it is known nowadays has
played an important role in the investigation of various issues
in general relativity~\cite{ksmh}.

The algebraic classification of the Ricci part,
known as the Segre classification, has been discussed by
several authors~\cite{Hall} and is of great importance in
three contexts. One is in understanding some purely geometrical
features of space-times~\cite{Churc}~--~\cite{BonaCollMorales1992}.
The second one is in classifying and interpreting matter
field distributions~\cite{Hall2}~--~\cite{SanRebTei}.
The third is as part of the procedure for checking whether apparently
different space-times are in fact locally the same up to coordinate
transformations (the equivalence problem~\cite{Cartan}~--~\cite{MM}).

In a recent article Santos {\em et al.\/}~\cite{SRT1}  have studied
the algebraic classification of second order symmetric tensors
defined on 5-dimensional (5-D for short) Lorentz
manifolds $M$, extending previous results
for 3-D and 4-D space-times~\cite{Hall3}~--~\cite{Hall5}.
Their analysis is made from first principles, i.e., without
using the previous classifications on lower dimensional space-times.
However, as concerns the classification itself
their approach is not straightforward.

In this work we examine the algebraic structure of
second order symmetric tensors defined on $5$-dimensional
space-times under different bases.
We shall assume the algebraic classification of the
Ricci tensor for 4-D space-times, and then show that the
algebraic classification Santos {\em et al.\/}~\cite{SRT1}
have studied can be achieved in a considerably simpler way.
Their results are, therefore, recovered under different premises.
We also state a fair number of new results concerning
the algebraic structure of the Ricci tensor and collect them
together in a theorem, which generalizes theorems on
4-D space-times~\cite{Hall,CorHal,Hall3}.
\vspace{4mm}

{\raggedright
\section{Prerequisites}   }  \label{pre}
\setcounter{equation}{0}

In this section we briefly review some important results required
for the remainder of this work. For a detailed account of the
notation used throughout this article see~\cite{Hall,SRT1}.

The classification of a second order symmetric real tensor at points
on a $5$-dimensional manifold carrying a Lorentz metric of signature
$\rm{(}-++++\rm{)}$ can be obtained directly from the analogous
classificationin the $4$-dimensional case where the results are
known~\cite{petrov2,Hall,Pleban,Hall3}. The situation in
the latter case can now be summarized in the following
theorem in which ${\bf t}, {\bf x}, {\bf y},{\bf z}$ and
${\bf l}, {\bf m}, {\bf x}, {\bf y}$ denote, respectively, a
pseudo-orthonormal and a null tetrad and in which the only
non-vanishing inner products are $ -t_at^a=x_ax^a=y_ay^a=z_az^a=1$
and $l_am^a=x_ax^a=y_ay^a=1$, respectively.
\vspace{2mm}

\begin{theorem}   \label{Segre4}
Let $(M,g)$ be a space-time so that $M$ is a $4$-dimensional manifold
and the metric $g$ has Lorentz signature $\rm{(} - + + + \rm{)}$.  Let
$p \in M$ and let $T$ be a real symmetric second order tensor at
$p$. Then (with respect to $g$) $T$ must take one of the Segre types
$\{1,111\}$, $\{211\}$, $\{31\}$ and $\{z\,\bar{z}11\}$ or a
degeneracy thereof.
The corresponding canonical forms in some appropriate tetrad are:
\begin{enumerate}
\item
for the type $\{1,111\}$ either of
\begin{eqnarray}
T_{ab} & = & 2\,\alpha_1 \, l_{(a} \, m_{b)} + \alpha_2 \, (l_al_b + m_am_b)
+ \alpha_3 \, x_ax_b + \alpha_4 \, y_ay_b \,,  \label{tab1111} \\
T_{ab} & = &(\alpha_2 - \alpha_1)\,t_at_b + (\alpha_1+\alpha_2)\,z_az_b
+ \alpha_3 \, x_ax_b + \alpha_4 \, y_ay_b\,,  \label{tab1111t}
\end{eqnarray}
\item
for the type $\{211\}$
\begin{equation}
T_{ab}  =  2\,\beta_1 \, l_{(a} \, m_{b)} \pm l_al_b
+ \beta_2 \, x_ax_b + \beta_3 \, y_ay_b\,,  \label{tab211}
\end{equation}
\item
for the type $\{31\}$
\begin{equation}
T_{ab}  =  2\,\gamma_1\,l_{(a} \, m_{b)} + 2\, l_{(a}\ x_{b)}
+ \gamma_1 \, x_ax_b + \gamma_2 \, y_ay_b\,, \label{tab31}
\end{equation}
\item
finally for the type $\{z\,\bar{z}11\}$
\begin{equation}
T_{ab}  =  2\, \delta_1 \, l_{(a} \, m_{b)} + \delta_2\, (l_al_b-m_am_b)
 + \delta_3\, x_ax_b + \delta_4 \, y_ay_b \,, \label{tabzz11}
\end{equation}
\end{enumerate}
where $\alpha_1,\alpha_2,\cdots ,\delta_4 \in  \R$  and
$\delta_2 \not= 0$.
\end{theorem}

{\raggedright
\section{The Algebraic Classification}   }  \label{class}
\setcounter{equation}{0}

Now suppose $M$ is a $5$-dimensional manifold carrying a Lorentz
metric $g$ and, with $p\in M$, let $T$ be a real second order,
symmetric tensor at $p$ with components $T^{ab}$ in some basis of
the tangent space $T_pM$ to $M$ at $p$. First note that any non-trivial
proper subspace of $T_pM$ may be classified into one of the three types,
{\em spacelike, null} and {\em timelike} according as it contains no null
directions, a unique null direction or more than one null direction,
respectively. Further, the $(g)$-orthogonal complement of a spacelike
(timelike) subspace is timelike (spacelike). Next, if the tensor
$T$, when regarded as a linear transformation from $T_pM$ to itself
with matrix $T^a_{\ b}$, admits a non-null invariant
subspace $V$ of $T_pM$ then the orthogonal complement of
$V$ is also an invariant subspace of $T$ as is easily checked
(see~\cite{Hall} for the case when $(M,g)$ is a space-time).
Now the tensor $T$ {\em must} admit an invariant $2$-dimensional subspace.
To see this note that if all the eigenvalues of $T$ are real then the
first two members (in the conventional numbering) of a canonical
Jordan basis span such a subspace whilst if $T$ admits a complex
eigenvalue the real and imaginary parts of any corresponding
complex eigenvector span an invariant subspace for $T$.
\vspace{3mm}

\begin{lemma}   \label{egvlema}
In the above notation $T$ must admit a spacelike eigenvector.
\end{lemma}
{\bf Proof} \ \
Suppose first that $T$ has a complex eigenvalue $\lambda =a+ib$ ($ a,b
\in \R\,, \ b\neq 0$) and corresponding complex eigenvectors
${\bf k} = {\bf r} +i\, {\bf s}$ \  (${\bf r}, {\bf s} \in
T_pM\,, \ {\bf s} \neq 0$). Then
\begin{equation}    \label{eng}
T_{ab}\,(r^b+is^b)=(a+ib)(r_a+is_a)\ .
\end{equation}
By replacing ${\bf k}$ with  $z {\bf k} $  ($z\in \C$),
if necessary, one can arrange that $s_ar^a=0$. On contracting
the real part of (\ref{eng}) with $s^a$ and the imaginary part
with $r^a$ one sees that $r_ar^a+s_as^a=0$. Now ${\bf r}$ and
${\bf s}$ cannot both be null for then they would be parallel
and (\ref{eng}) would give $b=0$. Hence one of ${\bf r}$ and
${\bf s}$ is timelike and the other spacelike and so the invariant
subspace they span is timelike.
The orthogonal complement of this subspace is thus spacelike and
the restriction of $T$ to it is a standard self adjoint Euclidean
action and is diagonalisable. It follows that $T$ has a spacelike
eigenvector lying in this orthogonal complement.

\begin{sloppypar}
Now suppose $T$ has only real eigenvalues. If $T$ admits a spacelike
invariant $2$-dimensional subspace the above argument applies, whilst
if $T$ admits a timelike invariant $2$-dimensional subspace the above
argument applies to its (invariant) orthogonal complement. So suppose
that $T$ admits only null invariant $2$-dimensional subspaces. The
unique null direction in each such null invariant subspace is, in fact,
an eigendirection of $T$ (see~\cite{Hall} for a proof in the
$4$-dimensional space-time case).
Let ${\bf k}_1$ be such a null eigendirection of $T$ with
eigenvalue $\mu_1\in \R$. If $T$ admits any other (independent)
eigenvector ${\bf k}_2$ with eigenvalue $\mu_2$ then ${\bf k}_2$
and $\mu_2$ are real since, by assumption, $T$ admits only real
eigenvalues. But then either $\mu_1=\mu_2$ (in which case the
subspace of $T_pM$ spanned by ${\bf k}_1$ and ${\bf k}_2$
consists entirely of eigenvectors and must contain a spacelike
member) or else $\mu_1\neq \mu_2$, in which case ${\bf k}_1$
and ${\bf k}_2$ are orthogonal (since $T$ is symmetric) and so
${\bf k}_2$ is spacelike, since ${\bf k}_1$ is null.
\end{sloppypar}

Thus the only remaining possibility is when $T$ has the single
(independent) eigenvector ${\bf k}_1$ which is real and null.
The Segre type for $T$ is then $\{5\}$ and its Jordan form  in
some (Jordan) basis at $p$ is
\[
T^a_{\ b}=\left[
\begin{array}{ccccc}
\mu_1 &   1   &   0   &   0   &   0   \\
  0   & \mu_1 &   1   &   0   &   0   \\
  0   &   0   & \mu_1 &   1   &   0   \\
  0   &   0   &   0   & \mu_1 &   1   \\
  0   &   0   &   0   &  0    &  \mu_1
\end{array}
\right]\,.
\]
The condition that $g_{ac}T^c_{\ b}$ is symmetric is then easily shown to
imply that $g_{12}=g_{22}=0$ at $p$. Thus the first two (column) vectors
in the Jordan basis are null and orthogonal at $p$, contradicting the
Lorentz signature of $g$, and so this case cannot occur.
This completes the proof.

\begin{theorem} \label{Segre5}
Let $M$ be a $5$-dimensional manifold admitting a Lorentz metric $g$
with signature $(-++++)$. Let $p\in M$ and let $T_{ab}$ be a real
symmetric second order tensor at $p$. Then (with respect to $g$) $T$
must take one of the Segre types $\{1,1111\}$, $\{2111\}$, $\{311\}$
and $\{z\,\bar{z}111\}$ or a degeneracy thereof. The corresponding
canonical forms in some appropriate pentad (either orthonormal
$\:{\bf t}, {\bf x}, {\bf y}, {\bf z}, {\bf w}\,$  or null
$\:{\bf l}, {\bf m}, {\bf x}, {\bf y}, {\bf w}\,$  in which
the only non-vanishing inner products are, respectively,
$-t_at^a=x_ax^a=y_ay^a=z_az^a=w_aw^a=1$ and
$l_am^a=x_ax^a=y_ay^a=w_aw^a=1$) \  at $p$ are:
\begin{enumerate}
\item
for the type $\{1,1111\}$ either of
\begin{eqnarray}    \label{tab11111}
T_{ab} & = & 2\,\alpha_1 \, l_{(a} \, m_{b)} + \alpha_2 \, (l_al_b + m_am_b)
+ \alpha_3 \, x_ax_b + \alpha_4 \, y_ay_b + \alpha_5 \, w_aw_b \,, \\
T_{ab} & = &(\alpha_2 - \alpha_1)\,t_at_b + (\alpha_1+\alpha_2)\,z_az_b \,,
+ \alpha_3 \, x_ax_b + \alpha_4 \, y_ay_b + \alpha_5 \, w_aw_b \,,
\end{eqnarray}
\item
for the type $\{2111\}$
\begin{equation}        \label{tab2111}
T_{ab}  =  2\,\beta_1 \, l_{(a} \, m_{b)} \pm l_al_b
+ \beta_2 \, x_ax_b + \beta_3 \, y_ay_b + \beta_4 \, w_aw_b \,,
\end{equation}
\item
for the type $\{311\}$
\begin{equation}        \label{tab311}
T_{ab}  =  2\,\gamma_1\,l_{(a} \, m_{b)} + 2\, l_{(a}\ x_{b)}
+ \gamma_1 \, x_ax_b + \gamma_2 \, y_ay_b + \gamma_3 \, w_aw_b \,,
\end{equation}
\item
finally for the type $\{z\,\bar{z}111\}$
\begin{equation}     \label{tabzz111}
T_{ab}  =  2\, \delta_1 \, l_{(a} \, m_{b)} + \delta_2\, (l_al_b-m_am_b)
+ \delta_3\, x_ax_b + \delta_4 \, y_ay_b + \delta_5 \, w_aw_b \,,
\end{equation}
\end{enumerate}
where $\alpha_1,\alpha_2,\cdots ,\delta_5 \in  \R$  and
$\delta_2 \not= 0$.
\end{theorem}

{\bf Proof} \ \
By the lemma $T$ admits a spacelike eigenvector, say ${\bf w}$
at $p$ scaled so that $w_a\,w^a=1$. The orthogonal complement
of ${\bf w}$ in $T_pM$ is a $4$-dimensional subspace $V$ of
$T_pM$ which is timelike (i.e. it admits the restriction of
$g(p)$ as a Lorentz metric) and invariant under $T$ and so
$T$ has a self adjoint action on $V$. So the restriction of
$T$ to $V$ is controlled by the previous theorem and, accordingly,
some appropriate tetrad may be chosen in $V$ to achieve
one of the canonical forms (\ref{tab1111}) -- (\ref{tabzz11}) for this
restriction. Appending the vector ${\bf w}$ to these tetrads
yields pentads in which $T$ takes the canonical forms
(\ref{tab11111})~--~(\ref{tabzz111}).

\vspace{3mm}

{\raggedright
\section{Further Results and Concluding Remarks}   }  \label{conclu}
\setcounter{equation}{0}

In this section we shall present some results concerning the
algebraic structure of a second order symmetric tensor $T$
defined on a $5$-dimensional Lorentz manifold $M$, which
generalizes theorems on $4$-dimensional space-times~%
\cite{Hall,Pleban,CorHal,Hall3}
and which can be collected in the following theorem:

\vspace{2mm}
\begin{theorem} \label{tabtheo}
Let $M$ be a real 5-dimensional manifold endowed
with a Lorentz metric $g$ of signature $\rm{(} - + + + + \rm{)}$.
Let $T$ be a real second order symmetric tensor
defined at a point $p \in M$. Then
\begin{enumerate}
\item
$T$ has a timelike eigenvector if and only if it
is diagonalizable over $\R$ (type \{1,1111\}).
\item
$T$ has at least three real orthogonal independent eigenvectors,
two of which (at least) are spacelike.
\item
$T$ has all eigenvalues real and is not diagonalizable
if and only if it has an unique null eigendirection.
\item
If $T$ has two linearly independent null eigenvectors
then it is diagonalizable over $\R$ (type \{1,1111\}).
\item
There always exists a 2-D spacelike subspace of $T_pM$
invariant under $T$.
\item
If a non-null subspace of $T_pM$ is invariant under
$T$, then so is its  orthogonal complement.
\item
There always exists a 3-D timelike subspace of $T_pM$ invariant
under $T$.
\item
If $T$ admits a null invariant subspace ${\cal N}$ of any
dimension $n$ {\rm(}=2, 3, or 4{\rm )} then it admits one
of each dimension 2, 3 and 4, and this occurs if and only if
$T$ admits a null eigenvector (which necessarily lies in any
such ${\cal N}$ admitted).
\end{enumerate}
\end{theorem}

The proofs can be gathered essentially by inspection of the canonical
forms (\ref{tab11111})~--~(\ref{tabzz111}) and are not
presented here for the sake of brevity (for proofs of similar
theorems in $4$-dimensional space-times see~\cite{Hall}).

To conclude, we remark that by a similar procedure to that
used in the lemma \ref{egvlema} one can show that a symmetric
two-tensor $T$ defined on an $n$-dimensional ($n \geq 4$)
Lorentz space has at least one real {\em spacelike\/}
eigenvector. The existence of this eigenvector can be used
to reduce, by induction, the classification of symmetric two-tensors
on $n$-dimensional ($n \geq 4$) spaces to the classification
on 4-dimensional spaces, thus recovering in a simpler way
the results of~\cite{SRT2}.


\vspace{5mm}

\begin{thebibliography}{99}

\bibitem{petrov1} A. Z. Petrov, {\em Sci. Notices of Kazan State
University\/} {\bf 114}, 55 (1954).

\bibitem{petrov2} A. Z. Petrov, ``Einstein Spaces'', Pergamon
Press (1969).

\bibitem{penrose1} R. Penrose, {\em Ann.\ Phys.\ NY\/} {\bf 10},
171 (1960).

\bibitem{penrose2} R. Penrose and W. Rindler , ``Spinors and
Space-Time'', vol.\ 2, Cambridge U. P., Cambridge (1986).

\bibitem{ksmh} See, for example, D. Kramer, H. Stephani, M. MacCallum
and E. Herlt, ``Exact Solutions of Einstein's Equations'', Cambridge
U. P., Cambridge (1980).

\bibitem{Hall} G. S. Hall, {\em Diff.\ Geom.\/} {\bf 12}, 53 (1984).
This reference contains an extensive bibliography on the classification
of the Ricci tensor on 4-dimensional space-times.

\bibitem{Churc} R. V. Churchill, {\em Trans.\ Amer.\ Math.\ Soc.\/}
{\bf 34}, 784 (1932).

\bibitem{Pleban} J. Pleba\'nski, {\em Acta Phys.\ Pol.\/} {\bf 26},
963 (1964).

\bibitem{CorHal} W. J. Cormack and G. S. Hall, {\em J. Phys.\ A\/}
{\bf 12}, 55 (1979).

\bibitem{BonaCollMorales1992} C. Bona, B. Coll and J. A. Morales,
{\em J. Math.\ Phys.\/} {\bf 33}, 670 (1992).

\bibitem{Hall2} G. S. Hall and D. A. Negm, {\em Int.\ J. Theor.\ Phys.\/}
{\bf 25}, 405 (1986).

\bibitem{Hall1} G. S. Hall, {\em Arab.\ J. Sci.\ Eng.\/} {\bf 9},
87 (1984).

\bibitem{FMM} J. J. Ferrando, J.A. Morales and M. Portilla,
{\em Gen.\ Rel.\ Grav.\/} {\bf 22}, 1021 (1990).

\bibitem{RebTei} M. J. Rebou\c{c}as and A. F. F. Teixeira,
{\em J. Math.\ Phys.\/} {\bf 32}, 1861 (1991).

\bibitem{RebTei1} M. J. Rebou\c{c}as and A. F. F. Teixeira,
{\em J. Math.\ Phys.\/} {\bf 33}, 2855 (1992).

\bibitem{SanRebTei} J. Santos, M. J. Rebou\c{c}as and A. F. F. Teixeira,
{\em J. Math.\ Phys.\/} {\bf 34}, 186 (1993).

\bibitem{Cartan} E. Cartan, ``Le\c{c}ons sur la G\'{e}om\'{e}trie des
\'{E}spaces de Riemann'', Gauthier-Villars, Paris (1951). Reprinted,
\'Edi\-tions Jac\-ques Ga\-bay, Pa\-ris (1988). English translation by J.
Glazebrook, Math.\ Sci.\ Press, Brookline (1983).

\bibitem{Karlh} A. Karlhede, {\em Gen.\ Rel.\ Grav.\/} {\bf 12},
693 (1980).

\bibitem{MacCSkea}
See M. A. H. MacCallum and J. E. F. Skea, ``{\sc sheep}:
A Computer Algebra System for General Relativity'', in {\em Algebraic
Computing in General Relativity, Lecture Notes from the First Brazilian
School on Computer Algebra\/}, Vol.\ II, edited by M. J. Rebou\c{c}as and
W. L. Roque.  Oxford U. P., Oxford (1994); and references therein.

\bibitem{Mac1} M. A. H. MacCallum, ``Classifying Metrics in Theory and
 Practice'', in {\em Unified Field Theory in More Than 4 Dimensions,
 Including Exact Solutions}, edited by V. de Sabbata and E. Schmutzer.
 World Scientific Publishing Co., Singapore (1983).

\bibitem{Mac2} M. A. H. MacCallum, ``Algebraic Computing in General
 Relativity'', in {\em Classical General Relativity}, edited by W. B. Bonnor,
 J. N. Islam and M. A. H. MacCallum. Cambridge U. P., Cambridge (1984).

\bibitem{MM} M. A. H. MacCallum, ``Computer-aided Classification of
Exact Solutions in General Relativity'', in {\em General Relativity
and Gravitational Physics (9th Italian Conference)\/}, edited by
R. Cianci, R. de Ritis, M. Francaviglia, G. Marmo, C. Rubano and
P. Scudellaro. World Scientific Publishing Co., Singapore (1991).

\bibitem{SRT1} J. Santos, M. J. Rebou\c cas and A. F. F. Teixeira,
J. Math.\ Phys.\ {\bf 36}, 3074 (1995).

\bibitem{Hall3} G. S. Hall, {\em J. Phys.\ A\/} {\bf 9}, 541 (1976).

\bibitem{Hall4} G. S. Hall, T. Morgan and Z. Perj\'es,
{\it Gen.\ Rel.\ Grav.\/} {\bf 19}, 1137 (1987).

\bibitem{Hall5} G. S. Hall, ``Physical and Geometrical
Classification in General Relativity'', Brazilian Centre for
Physics Research Monograph, CBPF-MO-001/93 (1993).

\bibitem{SRT2} J. Santos, M. J. Rebou\c cas and A. F. F. Teixeira,
{\em Gen.\ Rel.\ Grav.\/} {\bf 27}, 989 (1995).

\end{thebibliography}
\end{document}